\documentclass[a4paper,12pt,english,german]{article}

\begin{document}

\centerline{\bf  QUANTUM STRUCTURES OF THE HYDROGEN ATOM }

\bigskip

J. Jekni\' c-Dugi\' c$^{\ast}$\footnote{Email:
jjeknic@pmf.ni.ac.rs}, M. Dugi\' c$^{\dag}$, A. Francom$^{\$}$, M.
Arsenijevi\' c$^{\dag}$

$^{\ast}${Department of Physics, Faculty of Science, Ni\v s,
Serbia}

$^{\dag}${Department of Physics, Faculty of Science, Kragujevac,
Serbia}

$^{\$}$Austin, TX 78748, USA

\bigskip

{\bf Abstract: } Modern quantum theory introduces quantum
structures (decompositions into subsystems) as a new discourse
that is not fully comparable with the classical-physics
counterpart. To this end, so-called Entanglement Relativity
appears as a corollary of the universally valid quantum mechanics
that can provide for a deeper and more elaborate description of
the composite quantum systems. In this paper we employ this new
concept to describe the hydrogen atom. We offer a consistent
picture of the hydrogen atom as an open quantum system that
naturally answers the following important questions: (a) how do
the so called "quantum jumps" in atomic excitation and
de-excitation occur? and (b) why does the classically and
seemingly artificial "center-of-mass + relative degrees of
freedom" structure appear as the primarily operable form in most
of the experimental reality of atoms?

\

\bigskip

{\bf I. INTRODUCTION}

\bigskip

W. H. Zurek$^{1}$ presents a criticism of the anachronistic
teaching approach of quantum mechanics which is that it appears to
be a disservice to the ideology required by modern degrees of
understanding.

"{\it Quantum mechanics has been to date, by and large, presented
in a manner that reflects its historical development. That is,
Bohr's planetary model of the atom is still often the point of
departure, Hamilton-Jacobi equations are used to ''derive'' the
Schr\" odinger equation, and an oversimplified version of the
quantum-classical relationship (attributed to Bohr, but generally
not doing justice to his much more sophisticated views) with the
correspondence principle, kinship of commutators and Poisson
brackets, the Ehrenfest theorem, some version of the Copenhagen
interpretation, and other evidence that quantum theory is really
not all that different from classical-especially when systems of
interest become macroscopic, and all one cares about are
averages-is presented.}"

This paper can be considered an attempt to remedy this shared
criticism by providing a more modern introduction that is less
encumbered by classical reasoning.

So to begin, the hydrogen atom is among the most and best
investigated of all physical systems serving as the paradigmatic
foundation of non-relativistic quantum physics and thus a wide
range of scientific and technological understanding.  Today the
science and theory has evolved and we are more interested in the
different "structures" also called "subsystems" of the Hydrogen
Atom, such as the Electron and Proton ($e+p$), as one kind
decomposition, and the Center-Of-Mass + Relative Position ($CM+R$)
as another kind of decomposition.

At first glance, the Electron and Proton are real things, and the
Center-Of-Mass and Relative Position are abstract ideas.  To the
modern view these structural decompositions are each no less real
than the other.  In classical physics, such structural or
subsystem considerations are generally considered artificial, that
is to say that a composite system's center-of-mass is usually an
empty point in space, not a physical object that can be observed
or realistically targeted in an experiment. However, in contrast
to this classical mindset, atomic physics phenomenology directly
considers this $CM+R$ atomic structure. This contrast is typically
ignored by virtually identifying the atomic $e$ and $R$ systems on
the one hand, and the $p$ and $CM$ systems on the other hand. cf.
Section II for details.

Entanglement Relativity$^{2-10}$ is an important, subtle, and as
yet often overlooked part of modern quantum theory which lends
itself directly to the clarification of these issues with
structures.  Therefore it is on the basis of Entanglement
Relativity$^{2-10}$ in Section III that we explore and highlight
these crucial subtleties in distinguishing between the two above
mentioned structures of the Hydrogen Atom and provide what we
consider to be a more modern and proper description of
experimental reality. By considering the Hydrogen Atom as an
open$^{11-14}$ quantum system in Section IV, we answer the
following important questions: (a) how do the so called "quantum
jumps" in atomic excitation and de-excitation occur? and (b) why
does the classical $CM+R$ structure appear as the primarily
operable form in most of the realistic experimental situations
with atoms in spite of its seemingly artificial nature. Section V
summarizes the answers obtained in a more general form. Section VI
is a Discussion and we conclude in Section VII.

\bigskip

{\bf II. THE HYDROGEN ATOM: THE STANDARD VIEW}

\bigskip

Historically, it was the observation of discrete energy spectra
for atoms of the different chemical elements which has been at the
core of the development of quantum theory. Physically, these
observations come from both the spectroscopic results as well as
from the inelastic collisions of the atoms with, for example,
electrons.

Detailed observations of the inelastic collisions between atoms
and electrons, and more precisely the loss of the total kinetic
energy in such collisions reveals that the internal energy of
atoms appears to be discrete, or "quantized", gaining or losing
only whole integer values of energy.  This is in sharp contrast
with the macroscopic world we live in wherein it is universally
observed that classical macroscopic bodies like billiard balls can
accelerate or be accelerated smoothly, that is to say with a
continuous energy spectra that is not quantized into discrete
integer-like jumps from one speed to another.   Further
investigations into the structure of matter has led to the
conclusion that the hydrogen atom is not a fundamental particle
itself but is instead a composition of a single electron and a
single proton ($e+p$), and this is now considered as the
fundamental non-relativistic model of the Hydrogen Atom.

In keeping with a classical-physics model such as the "Solar
System Model" for atoms, the atomic internal degrees of freedom
refer usually to the atomic electrons ($e$). This then has served
as the basis for commonly used thoughts and terms such as
"electron orbits" and "electron energies" in describing the
internal structure and internal energy of the atomic elements.
However, this is imprecise and as we emphasize in Section V,
physically naive.

\bigskip

{\bf A. THE TEXTBOOK THEORETICAL MODEL OF THE HYDROGEN ATOM}

\bigskip

Bearing in mind the fundamental definition $e+p$, the Hydrogen
Atom is defined by way of the following Hamiltonian where the
model is non-relativistic and spin is not considered

\begin{equation}
\hat H = \hat T_e + \hat T_p  + \hat V_{Coul} = {\hat {\vec
p}_e^2\over 2m_e} + {\hat {\vec p}_e^2\over 2m_p} - k {e^2 \over
\vert \hat {\vec r}_e - \hat {\vec r}_p\vert}.
\end{equation}

\noindent In Eq. (1): $k$ is the standard Coulomb constant, $k =
(4 \pi \epsilon_{\circ})^{-1}$, and the indices $e$ and $p$ refer
to the  electron and the proton, respectively.

However, Eq. (1) is {\it  not} really the standard
quantum-mechanical theory of the Hydrogen Atom. In keeping with
the formalism of classical mechanics, what is considered standard
here is the alternative form with separation of variables. These
new variables simplify the calculation as well as present an
alternative description:

\begin{equation}
\hat H = \hat T_{CM} + \hat H_R = {\hat {\vec P}_{CM}^2\over 2M} +
{\hat {\vec p}_R^2\over 2\mu_R} - k {e^2 \over \vert \hat {\vec
\rho}_R \vert}, M = m_e + m_p, \mu_R = (m_e^{-1} + m_p^{-1})^{-1}.
\end{equation}

\noindent In Eq. (2), the alternate degrees of freedom are the
atomic center-of-mass ($CM$) and the "relative position" ($R$),
and their formal presentation is given in Eq. (5) below.  It
follows from Eq. (2), that the formal systems $CM$ and $R$
simplify the model by separating the variables, and there is not
any coupling between the $CM$ and $R$ variables.

Physically, the degrees of freedom provided by $CM$ describe the
atom as a whole, very much like the classically Newtonian or
"ballistic" description.  The internal degrees of freedom and the
related internal atomic energies refer to the atomic system $R$.
Therefore and thereby we can present the solution of the Schr\"
odinger equation for $R$ in a form that is uncoupled from the
Schr\" odinger equation for $CM$:

\begin{equation}
\hat H_R \vert nlm_l \rangle_R = E_n \vert nlm_l \rangle_R
\end{equation}

\noindent Eq. (3) above gives both the eigenvalues ($E_n$) and the
eigenstates ($\vert nlm_l \rangle_R$) of the $R$'s Hamiltonian
$\hat H_R$; $n$ is the energy (the 'principal') quantum number,
while $l$ is the angular-momentum- and  $m_l$ the magnetic-quantum
number. This resolves the classically paradoxical discrete energy
spectra for the internal degrees of freedom for the chemical
elements. Every value of the principal quantum number $n=1,2,3,. .
.$ defines one possible internal energy of the atom. By taking the
spin into consideration (in the perturbative manner), the other
quantum numbers can also contribute to defining the possible
internal energies of the atom, while bearing in mind their
well-known relations, $l = 1, 2, ,3, . . ., n-1$  and $m = -l,
-l+1, . . ., 0, . . .,  l-1, l$. In the position-representation,
the internal-energy ($\hat H_R$) eigenstates $\vert n l m_l
\rangle_R$ obtain the familiar form of the "wave-function",
$\psi_{nlm_l} ({\vec \rho}_R)$.

Most tasks$^{15}$ that we are generally faced with while working
with the atoms deal with the seemingly non-fundamental and
seemingly abstract $CM+R$ structure.  For example, atomic and
molecular interferometry and manipulation rely heavily on the $CM$
system, while in atomic spectroscopy we refer fairly exclusively
to the $R$ system.  And so we might say that our everyday
experience and phenomenology rely practically and almost
exclusively on the $CM+R$ structure and form.

\bigskip

{\bf B. THE BASIC PHENOMENOLOGICAL RULE OF ATOMIC PHYSICS}

\bigskip

The success of atomic physics is intimately linked to the
following phenomenological rule:

\smallskip

{\it An isolated atom spontaneously decays by making a transition
from an excited state with higher energy $E_m$ into a lower
energy-state $E_n$ by emitting a photon of the frequency $\nu =
(E_m - E_n) / h$, where $h$ is the Planck's constant}.

\smallskip

These transitions which ultimately lead to the only stable state,
the so called 'ground-energy state', are commonly described in
terms using phrases such as "Quantum Jumps". "Quantum Jumps" is a
deprecated form of phrasing and reasoning that has proven
incompatible with the Schr\" odinger Law.

As it is well known, Eq, (3) is equivalent with the time evolution
governed by the unitary operator $\hat U (t) = \exp (-\imath t
\hat H_R/\hbar)$. Then the probability of transition $\vert n l
m_l \rangle_R \to \vert n' l' m'_{l'}\rangle_R$, $\vert _R\langle
n' l' m'_{l'} \vert \hat U (t) \vert n l m_l \rangle_R \vert^2 =
0$ in every instant in time, $t$, unless $n=n', l=l',
m_l=m'_{l'}$. Physically, "Quantum Jumps" require intervention of
an external field or system on the atom. The way out of this
apparent contradiction is offered by modern theory of open quantum
systems, as we express in Section IV.

In the next section we emphasize some subtleties regarding the
Hydrogen Atom's structure that prepare for the analysis presented
in Section IV.

\bigskip

{\bf III. QUANTUM STRUCTURES}

\bigskip

Symbolically, the Hydrogen Atom (HA) can be presented as: $e + p =
HA = CM + R$. In the quantum-mechanical formalism, it means that
the HA Hilbert state space, $\mathcal{H}$, can be factorized as
$\mathcal{H}_e \otimes \mathcal{H}_p$ likewise as
$\mathcal{H}_{CM} \otimes \mathcal{H}_R$. Of course,
$\mathcal{H}_e \otimes \mathcal{H}_p = \mathcal{H} =
\mathcal{H}_{CM} \otimes \mathcal{H}_R$. Bearing this in mind, the
HA Hamiltonian, $\hat H$, Eqs. (1)-(2), can be presented precisely
as:

\begin{equation}
\hat T_e \otimes \hat I_p + \hat I_e \otimes \hat T_p + \hat
V_{Coul} = \hat H = \hat T_{CM} \otimes \hat I_R + \hat I_{CM}
\otimes \hat H_R.
\end{equation}

\noindent In Eq. (4) appear the "identity operators" $\hat I$ for
the respective factor-spaces of $\mathcal{H}$.

The two structures of the Hydrogen Atom, $e + p$ and $CM + R$, are
mutually linked by the well-known linear (and therefore
invertible) canonical transformations (LCT) that introduce $CM$
and $R$ as follows:

\begin{equation}
\hat {\vec R}_{CM} = [m_e \hat {\vec r}_e + m_p \hat {\vec r}_p ]/
(m_e + m_p), \quad \hat {\vec \rho}_R = \hat {\vec r}_e -  \hat
{\vec r}_p.
\end{equation}

It is a classical and not quantum mechanical reasoning which
states that the Linear Canonical Transformations can serve only as
a mathematical tool or that they exist as a mathematical artifact,
and not yet as a physically relevant method referring in any way
to the physical reality of physical objects. Indeed, a pair of
apples' center-of-mass is an empty point in space, not an object.
However, this classical statement is in sharp contrast to the core
of atomic phenomenology, which refers mainly to the seemingly
non-fundamental atomic structure $CM+R$. Resolving this apparent
conflict requires the careful and thoughtful analysis presented in
the remainder of this paper.

\bigskip

{\bf A. ENTANGLEMENT IN THE HYDROGEN ATOM}

\bigskip

Entanglement Relativity (ER) is a recently established
quantum-mechanical {\it rule}. As a corollary of quantum
mechanics, ER establishes$^{2-10}$:

\smallskip

\noindent ($\mathcal{E}\mathcal{R}$)
 {\it Typically, a separable form of a pure quantum state}
$\vert \Psi\rangle$ {\it for one decomposition obtains the
entangled form for another decomposition of a composite system}.

\smallskip

In more picturesque terms, there is entanglement for every quantum
state of a composite system as this follows from kinematics and
dynamics.  Given these considerations, the model of the Hydrogen
Atom serves as paradigmatic.

--{\it Kinematic arguments}. Consider the product of two functions
$f(x) g(y)$ of two independent variables, $x$ and $y$. If we
introduce new variables defined as $\xi = (x+y)/2$ and $\eta = x-
y$, then equality applies, $f(x) g(y) = f((2\xi + \eta)/2) g((2\xi
- \eta)/2)$. The point is that typically the new functions
$F(\xi)$ and $G(\eta)$ do not exist, such that $f(x) g(y) = F(\xi)
G(\eta)$.

Getting back to the Hydrogen Atom, the atomic instantaneous state
for the $CM+R$ decomposition is of the separable tensor-product
form $\vert \chi \rangle_{CM} \vert n l m_l \rangle_R$. In the
position representation this form is of the type considered above,
$f(x) g(y)$ (cf. also Eq. (7) below). Now, introducing the $e$ and
$p$ degrees of freedom (the inverse to Eq. (5)), this separability
of the state's form is lost--the quantum state for the $e+p$
decomposition becomes entangled, that is, of the form $\sum_i c_i
\vert \phi_i \rangle_e \vert \varphi_i \rangle_p$.

The rigorous proof of this  finding can be found in the literature
e.g.$^{6,10}$.

--{\it Dynamic arguments}. Interacting quantum systems are always
entangled. So it is with the atomic $e+p$ structure. Actually, the
Coulomb interaction in Eq. (1) entangles the hydrogen-atom's
electron and proton. In an instant of time, the atomic state for
the $e+p$ structure obtains the entangled form, $\sum_i c_i \vert
\phi_i \rangle_e \vert \varphi_i \rangle_p$. On the other hand,
due to non-interaction of the atomic $CM$ and $R$ systems for the
isolated atom, the Schr\" odinger dynamics for the atom as a whole
preserves the tensor-product form $\vert \chi\rangle_{CM} \vert n
l m_l\rangle_R$ for every instant of time. So, the Schr\" odinger
dynamics for the hydrogen atom as a whole {\it simultaneously}
preserves separability of states for $CM+R$ and induces
entanglement for $e+p$ structure. A dynamical proof of the
presence of entanglement for the $e+p$ structure can be found in
the literature$^{16}$.

In summary so far, we emphasize that as the universally valid
quantum mechanics establishes a unique quantum state in every
instant in time, there is the following equality stemming from
both kinematic and dynamic considerations for the Hydrogen Atom:

\begin{equation}
\vert \chi \rangle_{CM} \otimes \vert n l m_l  \rangle_R = \sum_i
c_i \vert \phi_i \rangle_e \otimes \vert \varphi_i \rangle_p.
\end{equation}

In the position-representation Eq. (6) reads:

\begin{equation}
\chi({\vec R}_{CM}) \psi_{nlm_l}({\vec \rho}_R) = \sum_i c_i
\phi_i({\vec r}_e) \varphi_i({\vec r}_p).
\end{equation}

\noindent While the standard theory (Section II) focuses on the
left hand side of Eq. (7), that is to say it deals with the "wave
functions" $\chi({\vec R}_{CM})$ and $\psi_{nlm_l}({\vec
\rho}_R)$, modern quantum mechanics extends this standard picture.

\bigskip

{\bf B. MANIPULATING THE ATOMIC DEGREES OF FREEDOM}

\bigskip

Like the concept of entanglement in Section III.A, the concept of
{\it local system} is also {\it relative}$^{3,5}$. For the
hydrogen atom, the 'electron' $e$ and the measurements of its
variables are local {\it only} for the $e+p$ structure. Regarding
the alternative $CM+R$ structure, the measurements performed on
$e$ are partially composite, or in other words "collective". To
see this, just invert the expressions in Eq. (5):

\begin{equation}
\hat {\vec r}_e = \hat {\vec R}_{CM} + m_p \hat {\vec \rho}_R/M,
\hat {\vec r}_p = \hat {\vec R}_{CM} - m_e \hat {\vec \rho}_R/M.
\end{equation}

So for example, a measurement of the electron's position $\hat
{\vec r}_e$ is partially a measurement of positions of {\it both}
$CM$ and $R$. The "total" measurement of the positions of both
$CM$ and $R$ reveals the values of the positions of both $e$ and
$p$. This {\it relativity of local system} is a general feature of
the composite system's structures. Formally, locality of a
subsystem is distinguished by the appearance of the "identity
operator", $\hat I$, for a given structure.  For example, for the
electron's position observable, $\hat {\vec r}_e \otimes \hat
I_p$, or for the $CM$ energy, $\hat H_{CM} \otimes I_R$, Eq. (6)
and Eq. (7), respectively.

To illustrate the relativity of "structure" (local system) from
the {\it operational} perspective, we briefly examine some typical
experimental situations not only with atoms.

--{\it Atomic de-excitation.} As it is emphasized in Section II.B,
it is a general spectroscopic fact that the higher (internal)
atomic energy quickly decays and is accompanied by emission of a
photon of some frequency $\nu$. The frequencies of the emitted
photons are characteristic for every chemical element.
Operationally, the decay is "directly" observed by observing the
emitted photon and by measuring its frequency. Now, bearing in
mind the different structures of the Hydrogen Atom, i.e. Section
III.A, we can say: Observation of the photon reveals that the $R$
state change (decay) and (in idealized situation) represents
indirect quantum measurement of the $R$'s initial energy. For such
physical situations, the "fundamental" structure, $e+p$, is of no
use--the local subsystem of the Hydrogen Atom that is of interest
is $R$ and its quantum state $\psi_{nlm_l} ({\vec \rho}_R)$, not
the electron $e$ and its states $\phi_i(\vec r_e)$, Eq. (7).

\smallskip

--{\it Interference experiments.} In past decades, atomic
interferometry has attracted much attention, for one example, by
releasing a cloud of metastable atoms towards a screen with the
two open slits (an analogue of the famous Young two-slit
experiment in optics), one can observe the interference fringes
(for a review see e.g. Section 6 in Ref. 15)--the points of the
impact of the individual atoms on the final screen. Clearly, these
spots reveal the atomic $CM$'s position on the final screen
without making any intervention between the atomic source and the
final screen. Certainly, the local subsystem of interest is the
atomic $CM$ system, not the atomic nucleus.

\smallskip

--{\it The Stern-Gerlach-like experiments and generalizations.}
Application of some external field usually couples external
degrees of freedom (e.g. the atomic $CM$ system) with the internal
degrees of freedom, such as the atomic spin in the Stern-Gerlach
experiment. If the external field(s) also couple the $CM$ and $R$
systems, one can design new procedures for manipulating the $CM$
dynamic control.  For example, in the Stern-Gerlach-like
experiments with molecules$^{17,18}$, this opportunity is very
fruitful. Again, the operationally preferred structure is $CM+R$.

\smallskip

--{\it Laser cooling of atoms.} The remarkable experiments reveal
the possibility to cool down an atomic gas, see e.g. Section 7 in
Ref. 15. By properly applying a laser field to the gas, one can
induce the internal atomic (the $R$'s) state transitions depending
on the atomic velocity (on the atomic $CM$ velocity). So, by
externally (partially) controlling the laser-light absorption and
the photon emission, one can manipulate with the atomic $CM$
kinetic energy (velocity). In effect, one can obtain lower
temperature of the gas. Again, the structure of interest is the
atomic $CM+R$ structure.

\smallskip

--{\it The semi-classical atomic  orbits.} The proper action
exerted on excited lithium atoms can have the effect of producing
'Kepler-like orbits' {\it in} the lithium atoms$^{19}$ . In the
manner described in Section II, these "orbits" are described as
the "electron's orbits". For a very short interval of time, the
atom resembles the classical Rutherford atom resembling a solar
system. However, in the theoretical support and explanation of the
experiment, the use of the $R$'s wave functions, $\psi_{nlm_l}
({\vec \rho}_R)$, {\it not} of the electron's states,
$\phi_i(\vec{r}_e)$, is made.
 So, borrowing the
notation for HA: the preferred structure of the atom in this
situation is
 $CM+R$, not $e+p$. This experiment is a remarkable
confirmation of the more general theoretical considerations$^{20}$
of the direct accessibility of the $R$ system. Only in
classical-physics terms, i.e. in visualization analogous to the
Solar-system's composition, may one express the effect in terms of
the atomic electron. While this supports intuition, it is,
strictly speaking, not physically correct$^{20}$.

\smallskip

--{\it Investigating the "structure of the matter".} As in the
epoch Rutherford's experiment, one can target the composite
systems by energetic quantum particles or fields in order to
observe the deeper spatial structure of a composite system. As to
the atomic species, ionizing the atoms reveals the presence of the
electrons, while bombarding the atoms by the energetic particles
may reveal the atomic nucleus. So, in these physical situations,
the preferred structure is $e+p$ and $e$ and $p$ appear as
"directly" accessible systems local to this structure.

From the above considerations, we learn: every physical situation
distinguishes a preferred structure of a composite system and the
related "directly observable" local systems.

\bigskip

{\bf C. EQUALITY AND NON-EQUIVALENCE OF THE  STRUCTURES}

\bigskip

For an isolated quantum system, the universally valid quantum
mechanics does {\it not a priori} set a privileged structure. By
Zanardi's [2]:

"{\it Without further physical assumption, no partition has an
ontologically superior status with respect to any other.}"

This is a direct consequence of the universally valid quantum
mechanics. A composite system's Hamiltonian is unique and the
system's quantum state is also unique in every instant in time.
The state typically takes different forms for different
structures, Section III.A. But the general rules and logic for
describing the subsystems are the same for every structure.

This democratic view to the HA structures is not applicable
anymore, regarding the {\it predictions} for the two structures.
In this sense, the two structures are not mutually equivalent.
E.g., complete knowledge of the electron's state is in no sense
sufficient for description of the atomic $CM$ or $R$ systems, and
{\it vice versa}. The wave functions for $e$, $\phi_i(\vec{r}_e)$,
and $R$, $\psi_{nlm_l}(\vec{\rho}_R)$ cannot be even compared to
each other; and analogously for the $p$ and $CM$ subsystems.
Mathematically, they belong to different "probability spaces",
e.g., the integration  $\int \vert \psi ({\vec \rho}_R)\vert^2
d^3{\vec r}_p$ does not provide the probability density for $e$.
As emphasized above, only the state of the atom as a whole
provides the probability density for arbitrary observable of the
atom, i.e. of any of the structures.

These subtle notions equally refer to the possible structures of
arbitrary composite quantum systems. To this end, the hydrogen
atom, as the simplest possible composite system, is paradigmatic.

\bigskip

{\bf IV. HYDROGEN ATOM IS AN OPEN QUANTUM SYSTEM}

\bigskip

As emphasized in  Section II.B, the phenomenological fact that the
"excited" internal-energy states are not stable clearly
demonstrates that the model of the isolated hydrogen atom, Eqs.
(1)-(3), is not correct or is at least not complete.

The proper physical picture offered in modern quantum theory
sticks to the later. The hydrogen atom is assumed to interact with
another physical system, the so called quantum vacuum fluctuations
(QVF). The total system "Atom + QVF" is now supposed to be
described by the Schr\" odinger Law which makes the atom
non-describable by the Schr\" odinger Law. The atom is then said
to be "open"$^{11}$. The quantum vacuum fluctuation system
effectively monitors the internal atomic ($R$) state and provides
the {\it smooth} (unitary and even time-reversible) state change
of the total system "Atom + QVF". With emission of one photon, the
state change reads, e.g.:

\begin{equation}
\vert n=2, l =0, m_l=0 \rangle_R \otimes \vert 0 \rangle_{QVF} \to
\vert n=1, l =0, m_l=0 \rangle_R \otimes \vert 1 \rangle_{QVF}.
\end{equation}

\noindent Now, {\it by ignoring}  the $QVF$ system in Eq. (9), one
obtains impression of the "quantum jumps" of Section II.B: $\vert
n=2, l =0, m_l=0 \rangle_R  \to \vert n=1, l =0, m_l=0 \rangle_R
$.

\bigskip

{\bf A. ARGUMENTS FROM THE THEORY OF OPEN QUANTUM SYSTEMS}

\bigskip

From Eq. (9) and from Section III.C, we learn: the environment
$QVF$ {\it targets only}  the atomic $R$-system and thus makes a
choice of the preferred structure $CM+R$ of the atom. Only for the
$CM+R$ structure the external influence of $QVF$ is local, and not
for $e+p$.

This is exactly the structure that is both "directly" accessible
(as described in Section III.B) as well as usually described in
the quantum mechanics textbooks (cf. Section II).

The related mathematical details can be found in the literature
e.g.$^{11,21,22}$. These considerations are open to further
technical improvements and there are also some open issues related
to interpretation (e.g. whether or not the $QVF$ system is
'realistic' or not). Bearing in mind that this closes the
conceptual gap in the phenomenological description of Section
II.B, we will leave these technical and issues of interpretation
aside.

Now, the {\it direct accessibility} of the atomic $CM+R$ structure
is easily realized: the environment makes $CM+R$ directly
accessible to observation, while the atomic HA structure $e+p$ is
"hidden" as its observation requires specific methods and
procedures well known from the experimental investigation of the
structure of matter.

The complexity of the theoretical modelling of atomic decays is an
obstacle to a more elaborated operational use of these models.
This is the reason that the phenomenological rule of Section II.B
is still in wide use.  Further progress in the foundations of the
theory of open quantum systems can be expected to change this
attitude.

\bigskip

{\bf B. QUANTUM MECHANICAL LIMIT FOR THE HYDROGEN ATOM}

\bigskip

For some higher energies the hydrogen atom breaks (ionizes) into a
pair ($e, p$) where there are not the atomic (internal) bound
states.  But the above rule for the preferred structure remains
the same: now every particle ($e$ and $p$) polarizes the vacuum
and separately induces the $QVF$-state changes$^{11}$. In effect,
$QVF$ {\it monitors every particle separately} and distinguishes
the well-known picture of the freely moving electrically charged
particles, in reference to the structure $e+p$. This is the
familiar picture from electrodynamics, both quantum and classical.

Again, according to Sections II.A and III.B, for the separable
state for $e+p$, it is in principle possible to observe
entanglement for the alternative structure $CM+R$. However, as
this is as yet speculative from the operational point of view, we
will not herein elaborate on this possibility any further.

\bigskip

{\bf V. PREFERRED QUANTUM STRUCTURES: AN OUTLOOK}

\bigskip

Sections III and IV provide us with the following lessons.

{\it First}, for an isolated system, there is no argument and/or
criterion or prescription to choose a preferred structure
(decomposition into subsystems) of a composite system.

{\it Second}, not only operationally, but also from the more
fundamental (e.g. the decoherence$^{11-14}$) point of view: {\it
the choice of the preferred structure is made by the
composite-system's environment}.

{\it Third}, only when considering an atom as open system, we
obtain answers to the issues of both "quantum jumps" (Section
II.B) and to the phenomenologically preferred structures (Section
III and IV) of the chemical elements. Being physically
incomparable and information-theoretically separated from the
atomic$R$ system, the atomic electron's "orbits" and "energies"
cannot be inferred
 from the information provided solely by the $R$'s quantum states
$\psi_{nlm_l}({\vec \rho}_R)$.

\bigskip

{\bf VI. DISCUSSION}

\bigskip

Modern quantum mechanics investigates quantum systems that
interact with their environments. Being, in principle,
undescribable by the Schr\" odinger law, such systems are termed
"open". The theoretical basis is the so-called Theory of Open
Systems, cf. e.g. $^{11}$. Constant progress in our ability to
comprehend open-systems dynamics opens new avenues in resolving
some long-standing issues in the standard quantum theory of the
isolated ("closed") quantum systems.

The exploration of quantum structures as we have expressed
throughout this paper is relatively new to modern quantum theory,
and while the structural variations are as old as quantum theory
itself, only recent progress places it in proper context in modern
quantum mechanics.  To this end, the physical model of the
Hydrogen Atom is paradigmatic.  Not only does such structural
analysis help to provide a deeper understanding, but also supports
a more consistent and further simplified description of the
Hydrogen Atom which, in turn, serves as a guide for deeper and
more sophisticated descriptions of composite quantum systems.

\bigskip

{\bf VII. CONCLUSION}

\bigskip

Modern Quantum Theory extends and also deepens our understanding
of the quantum world.  The lessons provided by the Theory of Open
Quantum Systems rely heavily on the structure of composite
systems.  In this paper we make use of both, and offer a fresh
look into the quantum mechanics of the hydrogen atom.  We
emphasize that modern quantum theory naturally and clearly answer
two important questions--(a) how do the so called "quantum jumps"
in atomic excitation and de-excitation occur? and (b) why does the
classically and seemingly artificial "center-of-mass + relative
degrees of freedom" structure appear as the primarily operable
form in most of the experimental reality of atoms?--and also that
the hydrogen atom model itself is paradigmatic, and can be used as
a guide in describing certain features of composite quantum
systems.

\bigskip

{\bf ACKNOWLEDGEMENTS} JJD, MD and MA acknowledge financial
support form Ministry of Science Serbia grant no 171028.

\bigskip

{\bf References}

\bigskip

[1] M. Dugi\' c, "What is 'system': the arguments from the
decoherence theory" arXiv:quant-ph/9903037v1

[2] P. Zanardi,  "Virtual Quantum Subsystems" Phys. Rev. Lett.
{\bf 87}, 077901 (2001)

[3] M. Dugi\' c, J. Jekni\' c,  "What is 'System': Some
Decoherence-Theory Arguments" Int. J. Theor. Phys. {\bf 45},
2249-2259 (2006)

[4] E. Ciancio, P. Giorda, P. Zanardi, "Mode transformations and
entanglement relativity in bipartite Gaussian states" Phys. Lett.
A {\bf 354}, 274-280 (2006)

[5] M. Dugi\' c, J. Jekni\' c-Dugi\' c,  "What Is 'System': The
Information- Theoretic Arguments" Int. J. Theor. Phys. {\bf 47},
805-813 (2008)

[6] A. C. De la Torre et al,  "Entanglement for all quantum
states" Europ. J. Phys. {\bf 31}, 325-332 (2010)

[7] N. L. Harshman, S. Wickramasekara, "Galilean and Dynamical
Invariance of Entanglement in Particle Scattering" Phys. Rev.
Lett. {\bf 98}, 080406 (2007)

[8] J. Jekni\' c-Dugi\' c, M. Dugi\' c,  "Multiple
System-Decomposition Method for Avoiding Quantum Decoherence"
Chin. Phys. Lett. {\bf 25}, 371-374 (2008)

[9] M. O. Terra Cunha, J. A. Dunningham, V. Vedral,  "Entanglement
in single-particle systems" Proc. R. Soc. A {\bf 463}, 2277-2286
(2007)

[10] J. Jekni\' c-Dugi\' c, M. Dugic, A. Francom, "Quantum
Structures of Model-Universe: Questioning the Everett
Interpretation of Quantum Mechanics", arXiv:1109.6424v1 [quant-ph]

[11] H.-P. Breuer, F. Petruccione,  "The Theory of Open Quantum
Systems" (Clarendon Press, Oxford, 2002)

[12] A. Riv\' as, S. F. Huelga, "Open Quantum Systems: An
Introduction" (SpringerBriefs in Physics, Springer, Berlin, 2011)

[13] D. Giulini, E. Joos, C. Kiefer, J. Kupsch, I.-O. Stamatescu,
H. D. Zeh,  "Decoherence and the Appearance of a Classical World
in Quantum Theory" (Springer, Berlin, 1996)

[14] W. H. Zurek, "Decoherence, einselection, and the quantum
origins of the classical" Rev. Mod. Phys. {\bf 75}, 715-775 (2003)

[15] G. Fraser, Ed., "The New Physics for the twenty-first
century" (Cambridge University Press, Cambridge, 2006)

[16] P. Tommasini, E. Timmermans, A.F.R.D. Piza, "The hydrogen
atom as an entangled electron-proton system" Am. J. Phys. {\bf
66}, 881-885 (1998)

[17] Y. Li, C. Bruder, C. P. Sun, " Generalized Stern-Gerlach
Effect for Chiral Molecules" Phys. Rev. Lett. {\bf 99}, 130403
(2007)

[18] E. Gershnabel, M. Shapiro, I. Sh. Averbukh, " Stern-Gerlach
deflection of field-free aligned paramagnetic molecules"
arXiv:1107.3916v1 [physics.chem-ph]

[19] H. Maeda,  D. V. L. Norum, T. F. Gallagher, "Microwave
Manipulation of an Atomic Electron in a Classical Orbit" Science
{\bf 307}, 1757-1760 (2005)

[20] A. V. Rau, J. A. Dunningham, K. Burnett, "Measurement-Induced
Relative-Position Localization Through Entanglement" Science {\bf
301}, 1081-1084 (2002)

[21] R. Graham, M. Miyazaki, "Dynamical localization of atomic de
Broglie waves: The influence of spontaneous emission" Phys. Rev. A
{\bf 53}, 2683-2693 (1996).

[22] Z. Zhu, H. Yu, S. Lu, "Spontaneous excitation of an
accelerated hydrogen atom coupled with electromagnetic vacuum
fluctuations" Phys.Rev. {\bf D73} 107501 (2006)

\end{document}